\documentclass[12pt]{article}

\usepackage[dvips]{graphicx}
\usepackage[english]{babel}

\usepackage{amssymb,epsfig}
\usepackage{amsmath}

\usepackage{slashed}
\usepackage[active]{srcltx}
\usepackage{psfrag}

\newcommand{\ms}{\mskip 1.5mu}

\def \e  {\mathop{\rm e}\nolimits}

\newcommand{\beq}[1]{
\begin{equation}\label{#1}}

\newcommand{\eeq}{\end{equation}}

\newcommand{\bea}[1]{
\begin{eqnarray}\label{#1}}

\newcommand{\eea}{\end{eqnarray}}

\begin{document}


\begin{center}
\textbf{\LARGE 
        Conformal symmetry of the Lange-Neubert evolution equation}

\vspace*{1.6cm}

{\large V.M.~Braun$\ms{}^1$  and  A.N.~Manashov$\ms{}^{1,2}$}

\vspace*{0.4cm}

\textsl{%
$^1$ Institut f\"ur Theoretische Physik, Universit\"at Regensburg,
D-93040 Regensburg, Germany \\
$^2$ Department of Theoretical Physics,  St.-Petersburg 
University, 199034, St.-Petersburg, Russia}

\vspace*{0.8cm}

\textbf{Abstract}\\[10pt]
\parbox[t]{0.9\textwidth}{
The Lange-Neubert evolution equation describes the scale dependence of the wave function of
a meson built of an infinitely heavy quark and light antiquark at light-like separations,
which is the hydrogen atom problem of QCD. It has numerous applications to the studies of
$B$-meson decays.
We show that the kernel of this equation can be written in a remarkably compact form, as a logarithm of the
generator of special conformal transformation in the light-ray direction. This representation allows one to study
solutions of this equation in a very simple and mathematically consistent manner.
Generalizing this result, we show that all heavy-light evolution kernels that appear
in the renormalization of higher-twist $B$-meson distribution amplitudes can be written in the same form.
}
\end{center}

\vspace*{1cm}


{\large \bf 1.}~~Studies of heavy meson weak decays have been instrumental to uncover the flavor sector
of the Standard model and can be a gate to new physics at TeV scales, if it exists.
Considerable effort has been invested to understand the QCD dynamics of heavy meson decays
in the heavy quark limit.
The B-meson distribution amplitude (DA), first introduced in \cite{Szczepaniak:1990dt}, provides the key
nonperturbative input in the QCD factorization approach \cite{Beneke:1999br} for weak decays involving
light hadrons in the final state.

Following an established convention we define the B-meson DA as the renormalized matrix element of the
bilocal operator built of an effective heavy quark field $h_v(0)$ and a light antiquark
$\bar q(zn)$ at a light-like separation:
\begin{eqnarray}\label{defDA}
\langle 0|\bar q(zn)\!\not\!n
       [zn,0]\Gamma  h_v(0)|\bar B(v)\rangle &=&
  -\frac{i}{2}F(\mu)\,\mbox{\rm Tr}\left[\gamma_5\!\not\!n\Gamma
   P_+\right]\,\Phi_+(z,\mu)\,
\end{eqnarray}
with
\begin{equation}\label{WL}
 {}[zn,0] \equiv {\rm Pexp}\left[ig\int_0^1\!d\alpha\,n_\mu A^\mu(\alpha z n)\right].
\end{equation}
Here $v_\mu$ is the heavy quark velocity,
$n_\mu$ is the light-like vector, $n^2=0$, such that $n\cdot v=1$,
$P_+=\frac12(1+\not\!v)$ is the projector on upper components of the
heavy quark spinor, $\Gamma$ stands for an arbitrary Dirac structure,
$|\bar B(v)\rangle$ is the $\bar B$-meson state in the heavy quark
effective theory (HQET) and $F(\mu)$ is the decay constant in HQET,
which is used for normalization.
The effective heavy quark  can be related to the Wilson line through
the following equation~\cite{Korchemsky:1991zp}:
\begin{equation}
       \langle 0| h_v(0) |h,v\rangle = [0,-v\infty] = {\rm Pexp}\left[ig\int_{-\infty}^0\!d\alpha\,v_\mu A^\mu(\alpha v)\right]\,,
\end{equation}
so that the operator in Eq.~(\ref{defDA}) can be viewed as a single light antiquark attached to the Wilson line
with a cusp containing  one lightlike and one timelike segment.

The invariant function $\Phi_+(z,\mu)$ where $z$ is a real number
defines what is usually called the leading twist
B-meson DA in position space. Its Fourier transform is
\begin{eqnarray}\label{Fourier}
     \phi_+(k,\mu)&=&\frac{1}{2\pi}\int_{-\infty}^{\infty}\!dz \, \e^{ikz}\Phi_+(z-i0,\mu)\,,
\nonumber\\
     \Phi_+(z,\mu)&=&\int_0^\infty\!dk \, \e^{-ikz}\phi_+(k,\mu)\,,
\end{eqnarray}
where in the first equation the integration contour goes below the singularities
of $\Phi_+(z,\mu)$ that are located in the upper-half plane.
The parameter $\mu$ is the renormalization (factorization) scale. We tacitly imply
using dimensional regularization with modified minimum subtraction.

The scale dependence of the DA is driven by the
renormalization of the corresponding nonlocal operator
$$O_+(z)= \bar q(zn)\not\!\!n\,[zn,0]\Gamma\, h_v(0).$$ The corresponding
one-loop $Z$-factor was computed by Lange and Neubert (LN) \cite{Lange:2003ff},
giving rize to an evolution equation which is convenient to write, for our purposes,
as a renormalization group equation for the operator $O_+(z)$~\cite{Braun:2003wx,Knodlseder:2011gc}:
\beq{Nils1}
 \left(\mu\frac{\partial}{\partial\mu}+\beta(g) \frac{\partial}{\partial g}+\frac{\alpha_s C_F }{\pi}\mathcal{H}\right)
{O}_+(z,\mu)=0\,,
\eeq
where
\beq{Nils2}
{} [\mathcal{H}f](z)= \int_0^1\frac{d\alpha}{\alpha}\Big(f(z)-\bar\alpha f(\bar\alpha z)\Big)
+\ln(i\mu z)\,f(z)-\frac54\,f(z)\,, \qquad \bar\alpha \equiv 1-\alpha\,.
\eeq
This equation thus governs the scale dependence of the $B$-meson DA in position space, $\Phi_+(z,\mu)$. It is fully equivalent
to the original LN equation for the DA in momentum space, $\phi_+(k,\mu)$, as it is easy to show by Fourier transformation.

\vskip 5mm


{\large \bf 2.}~~We will demonstrate that the LN kernel (\ref{Nils2}) can be written
in terms of the generators of collinear conformal transformations
\beq{generators}
  S_+ = z^2 \partial_z + 2j z \,, \qquad S_0 = z\partial_z + j\,, \qquad S_- = -\partial_z\,,
\eeq
where $j=1$ is the conformal spin of the light quark.
They satisfy the standard $SL(2)$ commutation relations
\beq{SL2}
  [S_+,S_-] = 2 S_0\,, \qquad [S_0,S_\pm] = \pm S_\pm\,.
\eeq
The starting observation is that the integral operator  $\mathcal{H}$ (LN kernel)
can be written in a somewhat different form by studying its action on the test functions
$f(z) = z^p$, $z\partial_z f(z) = p f(z)$. Here and below $\partial_z = \partial/\partial z$. In this way one obtains
\beq{Nils2a}
{} [\mathcal{H}f](z)= \left[\psi(z\partial_z+2)-\psi(1)+\ln(i\mu z)-\frac54\right]f(z)\,.
\eeq
Next, we use the identity for a fractional derivative $(i\partial_z)^a$
defined as the multiplication operator $k^a$ in momentum representation~\cite{Belitsky:2014rba}:
\begin{align}
(i\partial_z)^a=(iz)^{-a}\frac{\Gamma(a-z\partial_z)}{\Gamma(-z\partial_z)}.
\end{align}
It holds for the functions $f(z)$ that are holomorphic in the lower complex half-plane ${\Im m}\, z <0$, $z \in \mathbb{C_-}$,
and vanish at infinity.
Fourier transform for such functions goes over positive momenta
$f(z) = \int_0^\infty dk \, e^{-ikz} \tilde{f}(k)$, $(i\partial_z)^a f(z) = \int_0^\infty dk \, e^{-ikz} k^a\,\tilde{f}(k)$,
corresponding in our case to positive values of the light-quark energy $\omega = k/2$ in the B-meson rest frame, cf. Eq.~(\ref{Fourier}).
Expanding this identity around $a=0$ one gets
\begin{align}
\ln(i\partial_z)=\psi(-z\partial_z)-\ln(iz)\,
\end{align}
and making an inversion $z\to -1/z$
\begin{align}
\ln(iz^2\partial_z)=\psi(z\partial_z)+\ln(iz)\,.
\end{align}
Finally, since for any function $f(z\partial_z) z = z f(z\partial_z+1)$, we can write this identity as
\begin{align}
z^{-2}\ln(iz^2\partial_z)z^{2}=\ln\big[i(z^2\partial_z+2z)\big]=\ln(iS^+)=\psi(z\partial_z+2)+\ln(iz)\,.
\end{align}
Comparing with Eq.~(\ref{Nils2}) we see that
\begin{align}
 {} \mathcal{H} = \ln(i\mu\,S^+) -\psi(1) -\frac54
\label{main}
\end{align}
which is our main result. Note that the scale $\mu$ under the logarithm is necessary
simply because $S_+$ has dimension $[\text{mass}]^{-1}$.

Alternatively, the same expression can be derived starting from the commutation relations for the LN kernel
obtained in Ref.~\cite{Knodlseder:2011gc}:
\beq{Nils3}
 {} [S_+,\mathcal{H}] =0\,, \qquad [S_0,\mathcal{H}] = 1\,.
\eeq
Since the problem has one degree of freedom --- the light-cone coordinate of the light quark ---
it follows from $[S_+,\mathcal{H}] =0$ that the operator $\mathcal{H}$ must be a \emph{function}
of $S_+$, $\mathcal{H} = h(S_+)$. This function can be found using the second commutation relation.
Let $S=S_0+1$. Then $S_+= z S$ and the relation $[S_0,h(S_+)] = 1$ can be written equivalently  as
$[S,h(z\,S)] = 1$. Taking into account that $[S, z\, S]= z\, S$ one obtains an equation on the function $h(s)$
\begin{align}
s\, h'(s)=1~\Longrightarrow~h(s)=\ln s+\text{constant}\,,
\end{align}
reproducing the result in Eq.~(\ref{main}) up to a (scheme-dependent) constant.

\vskip 5mm


{\large \bf 3.}~~The main advantage of  Eq.~(\ref{main}) is that diagonalization of the kernel $\mathcal{H}$ can be
traded for a much simpler task of diagonalization of the first-order differential operator $S_+$ (\ref{generators}).
Eigenfunctions of $S_+$ take a simple form%
\footnote{The sign is chosen such that $Q_s(z)$ are real and positive for $z=-i\tau$, $\tau>0$.}
\begin{align}\label{eigenS+}
Q_s(z)=-\frac{1}{z^2} e^{is/z}\,, && iS_+\,Q_s(z) =   s \,Q_s(z)\,,
\end{align}
so that
\begin{align}\label{eigenH}
\mathcal{H}\, Q_s(z) & = \Big[ \ln(\mu\,s) -\psi(1) -\frac54 \Big]\,Q_s(z)\,.
\end{align}
A further advantage is that one can use $SL(2)$ representation theory methods to work with these solutions,
see e.g. Ref.~\cite{Derkachov:2002tf} for a short discussion of this technique.
In particular one can make use of the standard $SL(2)$ invariant scalar product~\cite{Gelfand} (for spin $j=1$)
\begin{align}
\langle{\Phi|\Psi}\rangle=\frac1\pi\int_{\mathbb{C}_-} d^2 z\, \overline{\Phi(z)}\,\Psi(z)\,,
\end{align}
where the (two-dimensional) integration goes over the lower half-plane $\mathbb{C}_-$, ${\Im m}\, z <0$.
The generator $i S^+$ is self-adjoint w.r.t. this scalar product.
The eigenfunctions~(\ref{eigenS+}) are orthogonal to each other and form a complete set
\begin{align}
\langle{Q_{s'}|Q_s}\rangle= \frac1s\delta(s-s')\,,
&&
 \int_0^\infty ds \,s\, Q_s(z) \overline{Q_s(z')} = \frac{e^{-i\pi}}{(z-\bar z')^2}\,.
\end{align}
The function on the r.h.s. of the completeness relation is called reproducing kernel~\cite{Hall}. 
It acts as a unit operator so that 
for any function holomorphic in the lower half plane
\begin{align}
   \Psi(z) = \frac1\pi\int_{\mathbb{C}_-} d^2 z'\  \frac{e^{-i\pi}}{(z-\bar z')^2} \Psi(z')\,.
\end{align}
Hence the $B$-meson DA (\ref{defDA}) can be expanded as
\begin{align}
\Phi_+(z,\mu)=\int_0^\infty ds \,s\,\eta(s,\mu)\, Q_s(z) = - \frac1{z^2}\int_0^\infty ds \,s\,e^{is/z}\,\eta(s,\mu)\,,
&& \eta(s,\mu) =\langle{Q_s|\Phi}\rangle\,.
\end{align}
The integration goes over all possible eigenvalues of the step-up generator $S_+$ that corresponds to
special conformal transformations along the light-ray $n^\mu$.
This representation is very similar to the one suggested in Ref.~\cite{Bell:2013tfa}.

The scale-dependence of the coefficients $\eta(s,\mu)$ is governed by the renormalization-group equation
\begin{align}
\Big(\mu\frac{\partial}{\partial \mu}+\beta(\alpha_s)\frac{\partial}{\partial {\alpha_s}}
  +\Gamma_{cusp}(\alpha_s)\ln (\mu\,s/s_0)\Big)F(\mu)\eta(s,\mu)=0\,,
\end{align}
where $s_0= e^{5/4-\gamma_E}$
and
$
\Gamma_{cusp}(\alpha_s)=\frac{\alpha_s}{\pi} C_F+\ldots
$
is the cusp anomalous dimension~\cite{Polyakov:1980ca,Korchemsky:1987wg}.

The solution of this equation takes the form
\begin{align}
F(\mu)\,\eta(s,\mu)& =F(\mu_0)\,\eta(\xi,\mu_0)\times\exp\left\{-\int_{\mu_0}^{\mu} \frac{d\tau}{\tau}\,
\Gamma_{cusp}(\alpha_s(\tau))\,\ln (\tau\,s/s_0)
\right\}
\notag\\
&
= F(\mu_0)\,\eta(\xi,\mu_0) \left(\frac{\mu_0\,s}{s_0}\right)^{r(\mu)} B(\mu)\,,
\end{align}
where
\begin{align}
r(\mu)&=-\int_{\alpha(\mu_0)}^{\alpha(\mu)} \frac{d\alpha}{\beta(\alpha)}\,
\Gamma_{cusp}(\alpha_s) = 2 C_F/\beta_0\,\ln\left(\frac{\alpha(\mu)}{\alpha(\mu_0)}\right)+\ldots\,,
\notag\\
B(\mu)&=\exp\left\{-\int_{\alpha(\mu_0)}^{\alpha(\mu)} \frac{d\alpha}{\beta(\alpha)}\,
\Gamma_{cusp}(\alpha)\int_{\alpha(\mu_0)}^{\alpha}\frac{d\alpha'}{\beta(\alpha')}\right\}.
\end{align}
In practical applications the momentum (energy) representation for the $B$-meson DA $\phi_+(k,\mu)$
as defined in (\ref{Fourier}) is more convenient. This can be derived easily by observing that
exponential functions $e^{-ipz}$, $p>0$ are mutually orthogonal and form a complete set
w.r.t. the same scalar product
\begin{align}
 \langle e^{-ipz}|e^{-ip'z}\rangle &= \frac{1}{p}\delta(p-p')\,.
\label{ortho}
\end{align}
Hence
\begin{align}
   \Phi_+(z,\mu) = \int_0^\infty dp\,p\, e^{-ipz} \langle e^{-ipz}|\Phi_+(z,\mu)\rangle
=
 \int_0^\infty dp\,p\, e^{-ipz} \int_0^\infty ds \,s\,\eta(s,\mu)\,\langle e^{-ipz}|Q_s(z)\rangle
\end{align}
and therefore
\begin{align}
     \phi_+(k,\mu)& = \frac{1}{2\pi}\int_{-\infty}^{\infty}\!dz \, \e^{ikz}\Phi_+(z-i0,\mu)
= k \int_0^\infty ds \,s\,\eta(s,\mu)\,\langle e^{-ikz}|Q_s(z)\rangle\,.
\end{align}
Using
\begin{align}
 \langle e^{-ikz}|Q_s(z)\rangle =
\frac{1}{\sqrt{ks}} J_1(2\sqrt{ks})
\label{project}
\end{align}
we finally obtain
\begin{align}
   \phi_+(k,\mu)& = \int_0^\infty ds \, \sqrt{ks}\, J_1(2\sqrt{ks})\,\eta(s,\mu)\,,
\label{phi(k)}
\end{align}
where $J_1(x)$ is the Bessel function. The representation in Eq.~(\ref{phi(k)}) is equivalent to the one
suggested by Bell, Feldmann, Wang and Yip
in Ref.~\cite{Bell:2013tfa}, who noticed that the evolution equation is significantly simplified 
in this manner. In their notation, cf. second line in Eq.~(2.17),
$s\,\eta(s,\mu) \equiv \rho_+(1/s,\mu)$.

The orthogonality relation (\ref{ortho}) combined with the projection (\ref{project})
leads to a familiar relation for the Bessel functions
\begin{align}
\int_0^\infty ds\,J_1(2\sqrt{ps })\,J_1(2\sqrt{p's })=\delta(p-p')\,,
\end{align}
which can be used to invert Eq.~(\ref{phi(k)}) and express $\eta(s,\mu)$ in terms of $\phi_+(k,\mu)$.

Note that the representation in (\ref{main}) is valid for the evolution kernel in momentum space as well,
but the generator $S_+$ has to be taken in the adjoint representation
\begin{align}
 \mathcal{S}_+ &= i \big[k \partial^2_k + 2j\partial_k\big]\,,\qquad j=1\,.
\end{align}
The Bessel functions appearing in (\ref{project}), (\ref{phi(k)}) are eigenfunctions of $\mathcal{S}_+$,
indeed:
\begin{align}
  s \langle e^{-ikz}|Q_s(z)\rangle = \langle e^{-ikz}|i S_+ Q_s(z)\rangle = \langle iS_+ e^{-ikz}|Q_s(z)\rangle = 
  i \mathcal{S}_+ \langle e^{-ikz}|Q_s(z)\rangle\,. 
\end{align}

Of particular interest for the QCD description of $B$-decays is the value of the first
negative moment
\begin{align}
  \lambda_B^{-1}(\mu) &= \int_0^\infty \frac{dk}{k} \phi_+(k,\mu) = \int_0^\infty d\tau \, \Phi_+(-i\tau, \mu) =  \int_0^\infty ds\, \eta(s,\mu)\,.
\end{align}
As demonstrated in~\cite{Bell:2013tfa}, QCD factorization expressions for $B$ decay amplitudes
can conveniently be written in terms of $\eta(s,\mu)$ as well, so that we do not dwell on this topic here.

\vskip 5mm


{\large \bf 5.}~~The same representation can be derived for arbitrary two-particle heavy-light one-loop 
kernels that contribute to the evolution equations for higher-twist $B$-meson DAs~\cite{Knodlseder:2011gc}.
The difference to the leading twist is that the two-particle evolution equations are not closed: 
The two-particle, $2\to2$, kernels appear as parts of larger mixing matrices involving $2\to 3$ parton transitions,
however, $3\to 2$ transitions do not occur at the one-loop level.  

Explicit expressions for all $2\to2$ heavy-light kernels have been derived in Ref.~\cite{Knodlseder:2011gc},
see Sec.~3.2.
They can be written in terms of an integral operator
\begin{align}
[\mathcal{H}_{j}f](z)=\int_0^1\frac{d\alpha}\alpha\Big[f(z)-\bar\alpha^{2j-1}f(\bar \alpha z)\Big]+\ln(i\mu z) f(z)
- [\sigma_h+\sigma_\ell]f(z)\,,
\label{Hj}
\end{align}
where $j$ is the conformal spin of the light parton $\ell$ (quark or gluon) and the constants 
$\sigma_h=1/2$, $\sigma_{\rm quark} = 3/4$, $\sigma_{\rm gluon} = \beta_0/4N_c$ ($\beta_0= 11/3 N_c - 2/3 n_f$) are
related to the anomalous dimensions of the fields.
Conformal spin of a parton is defined as $j=(d+s)/2$ where $d$ is canonical dimension and $s$ is spin projection  on the 
light cone, see~\cite{Braun:2003rp}.
For a quark $j=1$ for the ``plus'' projection that contributes to the leading-twist $B$-meson DA
(\ref{defDA}), in which case (\ref{Hj}) reproduces (\ref{Nils2}), and $j=1/2$ for the ``minus'' projection
that is relevant for the DA $\Phi_-(z,\mu)$, cf.~\cite{Beneke:1999br}. 
In turn, for a gluon $j=3/2$ for the leading-twist projection and $j=1$ for the higher-twist.

Following the above derivation for $j=1$ we obtain the following representation for the kernel in the 
general case:
\begin{align}
\mathcal{H}_j = \ln(i\mu S^{(j)}_+)-\psi(1) - \sigma_h-\sigma_\ell\,,
\end{align}
where the generator of special conformal transformations $S^{(j)}_+$ for spin $j$ 
is defined in Eq.~(\ref{generators}).
The eigenfunctions of $S^{(j)}_+$ have the form
\begin{align}
Q_s^{(j)}(z)=\frac{e^{-i\pi j}}{z^{2j}} e^{is/z}\,, && iS^{(j)}_+\, Q_s^{(j)}(z)=s\,Q_s^{(j)}(z)\,.
\end{align}
They are orthogonal and form a complete set with respect to the $SL(2)$ scalar product~\cite{Hall}
\begin{align}
\langle\Phi|\Psi\rangle_j &=\frac{2j-1}\pi\int_{\mathbb{C}_-} \mathcal{D}_j z\,  \overline{\Phi(z)}\,\Psi(z)\,,
\end{align}
where $\mathcal{D}_j z=d^2z\, [i(z-\bar z)]^{2j-2}$. One obtains
\begin{align}
\langle{Q_s^{(j)}|Q_{s'}^{(j)}\rangle}_j=\frac{\Gamma(2j)}{s^{2j-1}}\,\delta(s-s')\,,
&&
\frac1{\Gamma(2j)}\int_0^\infty ds \, s^{2j-1}\,Q_s^{(j)}(z)\,\overline{Q_s^{(j)}(z')} = \frac{e^{-i\pi j}}{(z-\bar z')^{2j}}\,.
\label{RK}
\end{align}
The expression on the r.h.s. of the second integral defines the reproducing kernel for arbitrary spin $j$~\cite{Hall},
i.e. for arbitrary function (holomorphic in the lower plane)
\begin{align}
\Psi(z)&=\frac{2j-1}\pi\int_{\mathbb{C}_-} \mathcal{D}_j z\, \frac{e^{-i\pi j}}{(z-\bar z')^{2j}}\,\Psi(z)\,.
\end{align}

The functions $Q_s^{j}(z)$ diagonalize the renormalization group kernel
\begin{align}
  \mathcal{H}_j Q_s^{j}(z)&= \big[\ln (\mu s) - \psi(1) - \sigma_h-\sigma_\ell \big]Q_s^{j}(z)
\end{align}
so that it is natural to write matrix elements of generic heavy-light operators as an expansion
\begin{align}
\Phi_j(z,\mu)=\int_0^\infty ds\, s^{2j-1} \eta_j(s,\mu) \,Q_s^{(j)}(z)\,,
\end{align}
where $\Phi_j(z,\mu)$ is analogue of $\Phi_+(z,\mu)$ (\ref{defDA}).

The expansion coefficients $\phi_j(k,\mu)$ appearing in the Fourier transform
\begin{align}
\Phi_j(z,\mu)=\int_0^\infty dk \, e^{-ikz}\,\phi_j(k,\mu)
\end{align}
can be found making use of the following relations:  
\begin{align}
\langle{e^{-ikz}|e^{-ik'z}\rangle}_j & ={\Gamma(2j)}\,{k^{1-2j}}\,\delta(k-k')\,,
\notag\\[2mm]
\langle e^{-ikz}|Q_s^{(j)}\rangle_j & = \Gamma(2j)\,(ks)^{1/2-j}\, J_{2j-1}(2\sqrt{ks})\,.
\end{align}
In this way one obtains 
\begin{align}
\phi_j(p,\mu) &=\int_0^\infty ds\,\eta_j(s,\mu)\, (sp)^{j-1/2}\,J_{2j-1}(2\sqrt{ps})\,.
\end{align}
In particular for $j=1/2$ corresponding to the $B$-meson DA $\phi_-(k,\mu)$ \cite{Beneke:1999br}
the conformal expansion goes over Bessel functions $J_{0}(2\sqrt{ks})$ as compared to  
$J_{1}(2\sqrt{ks})$ for the leading twist, cf.~\cite{Bell:2013tfa}. 


\vskip5mm

{\large \bf 6.}~~To summarize, we have constructed a conformal expansion of the distribution amplitudes 
of heavy-light mesons in terms of eigenfunctions of the generator of special conformal transformations.
This construction is similar in spirit to the well-known expansion of DAs of light mesons in 
Gegenbauer polynomials which are eigenfunctions of two-particle $SL(2)$ Casimir operators, see e.g.~\cite{Braun:2003wx}. 
Similar to the latter case, this expansion can serve as a basis for the construction of approximations of 
phenomenological relevance.

As we have shown, this expansion is a consequence of the commutation relations (\ref{Nils3})
and it would be very interesting to find out whether these relations hold true to all orders in perturbation theory 
for a conformal theory like $N=4$ SYM. The consequences of our results for the DAs of baryons made of one heavy and 
two light quarks should be studied as well. 


\vskip5mm

{\large \bf 7.}~~{\large\bf Acknowledgements}\\
This work was supported by the DFG, grant BR2021/5-2.

\end{document}